\documentclass[manuscript]{acmart}

\usepackage{hyperref}
\AtBeginDocument{%
  \providecommand\BibTeX{{%
    Bib\TeX}}}

\setcopyright{acmlicensed}
\copyrightyear{2018}
\acmYear{2018}
\acmDOI{XXXXXXX.XXXXXXX}
\acmConference[Conference acronym 'XX]{Make sure to enter the correct
  conference title from your rights confirmation email}{June 03--05,
  2018}{Woodstock, NY}
\acmISBN{978-1-4503-XXXX-X/2018/06}

\usepackage{graphicx}
\usepackage{subcaption}
\usepackage{float}      
\usepackage{placeins}   

\usepackage{hyperref}

\begin{document}

\title{"Because we are no longer ashamed of our disabilities, we are proud": Advocating and Reclaiming Next-Gen Accessibility Symbols
}

\author{Karen Joy}
\affiliation{%
  \institution{Rutgers University}
  \state{New Jersey}
  \country{USA}
}
\email{kj461@scarletmail.rutgers.edu}

\author{Chris Dodge}
\affiliation{%
  \institution{Ipsos New York}
  \state{New York}
  \country{USA}
}

\author{Harsh Chavda}
\affiliation{%
  \institution{Purdue University}
  \state{Indiana}
  \country{USA}
}

\author{Alyssa Sheehan}
\affiliation{%
  \institution{Ipsos Atlanta}
  \state{Georgia}
  \country{USA}
}


\renewcommand{\shortauthors}{Joy et al.}

\begin{abstract}
  Our study investigates the relationship between accessibility symbols and emerging technologies in supporting disability disclosure. We conducted 23 remote co-creation sessions with semi-structured interviews to examine participants’ awareness of existing symbols, how they use symbols across online and offline contexts, and barriers to adoption and interpretation. Through participant sketching and future-oriented storyboard probes, participants proposed ways to integrate symbols into wearable devices, mobile interfaces, and portable tools, emphasizing customizable and context-sensitive disclosure. Our findings suggest symbols are most effective when paired with technologies that provide user control over visibility and optional pathways for explanation, helping reduce misinterpretation while supporting agency in disclosure moments. By reimagining symbol-based assistance as part of a broader disclosure system—where meaning depends on the symbol, its carrier, and context—this work informs more inclusive accessibility supports across diverse settings.
\end{abstract}




\received{20 February 2007}
\received[revised]{12 March 2009}
\received[accepted]{5 June 2009}

\maketitle

\section{Introduction}

The disclosure of disabilities has gained attention in HCI and design, particularly in the context of health and accessibility technologies. Researchers are exploring how individuals with invisible chronic illnesses \cite{1ammari, 2Workplace}, especially chronic invisible or “hidden disabilities” \cite{2Workplace}, share information about their conditions across various platforms \cite{3Sannon}. These works delve into individuals’ comfort with anonymous disclosures online or uncertainty about unavoidable disclosures, such as those in a public workplace \cite{designempathysocialmedia}. Our work builds on these studies of invisible disabilities and underscores the importance of designing for disclosure through community engagement and speculative storyboards. Most research explains why people disclose, but our research interrogates what and how they disclose.

Accessibility disclosure—the act of openly identifying one's disability status—varies significantly between online and physical environments \cite{26chen2011narrative,27nario2013redefining}. Disclosing any disability is complex and dynamic, shaped by internal, interpersonal, organizational, and societal factors \cite{4santuzzi2016}. Accessibility symbols, such as the International Symbol of Access (ISA) \footnote{\url{https://en.wikipedia.org/wiki/International_Symbol_of_Access}}, have long served as critical tools for disclosure, signaling inclusion and support for individuals with disabilities. 

However, these symbols often fail to meet the nuanced and evolving needs of the communities they aim to represent. The design of symbols, such as the wheelchair figure, have long been criticized for their limited scope, focusing narrowly on physical disabilities while excluding broader, less visible conditions \cite{guffey2}. Guffey explores the intricate challenges of designing universal accessibility symbols, highlighting how cultural and institutional pressures often favor simplicity and standardization over community-driven contributions \cite{guffey2}. This approach to designing accessibility symbols has been found to be counterproductive, inadvertently reinforcing stereotypes, marginalizing diverse experiences, and failing to capture the intersectionality of disability \cite{4santuzzi2016, 6Hofmann,20carroll2023designers}. 

\begin{figure}[t]
  \centering
  \includegraphics[width=\linewidth]{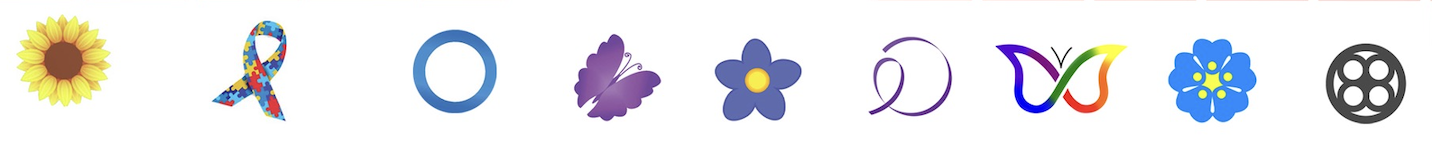}
  \Description{Different disability symbols.}
  \caption{Disability Symbols}
  \label{fig:symbols}
\end{figure}

Consequently, individuals frequently have limited opportunities to control or personalize these symbols, leading to disclosure experiences that may not fully align with their personal preferences and identities \cite{6Hofmann,31barstow2019examining}. In contrast, the rise of social media and online platforms has given birth to a new generation of accessibility symbols (See Figure.~\ref{fig:symbols}). The customization of digital content allows for more nuanced and personalized disclosures than traditional offline methods.  Symbols such as the Hidden Disabilities Sunflower, the Autism Puzzle Ribbon, and the Blue Diabetes Circle [7] have gained traction for their ability to represent specific conditions in more meaningful and personalized ways \cite{7das2021blue}. We even see these symbols in the ACM research community used as disability signifiers for participant populations \cite{8RixenConsent, 9Angerbauer}.

Understanding the evolving meanings and impacts of these symbols is crucial for creating truly inclusive environments, especially as we begin to see them transition into technologies such as augmented and virtual reality \cite{18Gualano}. These technologies enable individuals and communities to create and share their own narratives, challenging traditional gatekeepers of content creation \cite{34wang2024virtual}. This shift not only democratizes representation but also highlights the potential of digital spaces to foster broader awareness and understanding of the diverse realities of disability \cite{34wang2024virtual, VRrepresent}. Together, these emerging symbols signal a new era of identity representation, one that has the potential to promote inclusivity, adaptability, and the voices of those they aim to serve.

The symbols encompass a diverse range of disability and health-related needs, each uniquely designed to represent specific communities. The participants with whom we engaged reported that these symbols resonate with them, aligning with their identities and the communities to which they belong. While disability categories are nuanced and varied, spanning chronic and acute conditions across different ages and demographics, these symbols are distinct and, in some contexts, have become recognizable signifiers for particular invisible conditions \cite{35towards}. Individuals with these disabilities often choose to identify with these symbols to communicate their needs. However, the effectiveness of these symbols is limited by a general lack of awareness, which diminishes their ability to convey information discreetly. This lack of recognition may also lead to unintended stigma or unwanted attention, discouraging individuals from using symbols to communicate their needs.

Our research investigates how emerging technologies can support and facilitate customized communication and information sharing through these next-gen symbols of accessibility for people with disabilities. We centered our work around the following questions:

\begin{itemize}

\item RQ1: Awareness: What is the level of awareness and recognition of existing accessibility symbols within the community?

\item RQ2: Usage: How do individuals with disabilities utilize these symbols to represent their specific needs and preferences? 

\item RQ3: Representation: What breakdowns in recognition and interpretation do participants foresee for accessibility symbols across contexts?

\item RQ4: Reimagination: How do participants reimagine symbols when paired with different technological modalities (wearables, mobile, portable, XR)?

\item RQ5: Motivations or Barriers: What factors influence the adoption or rejection of these symbols by individuals with disabilities?

\end{itemize}

By addressing these questions, our research aims to inform the design of more inclusive and effective technologies that support personalized communication and disability disclosure. This paper contributes an empirical account of how people with invisible disabilities reimagine “next-gen” accessibility symbols when paired with emerging modalities (wearables, mobile, portable, XR), showing where symbols are most effective. Conceptually, we frame symbol-based assistance as a disclosure system—where meaning depends on the symbol, its carrier, and context—and use Peirce’s triad to explain misinterpretation and “interpretant" drift. \cite{19palinoan2024charles,20carroll2023designers, 22irvine2023semiotics}

We also reveal how symbols represent and shape individual identities using a semiotic lens. Semiotics is the study of signs and symbols and their interpretation. C. S. Peirce used the terms "sign," “symbol,” and "representament" interchangeably \cite{22irvine2023semiotics}. In semiotics, a sign encompasses the relationship between the sign and its object and how an interpreter assigns meaning to it. Notably, there is a lack of studies exploring the relationship between information and symbolic representation. The few studies that do examine their relationship suggest that symbols and icons conveying information should be viewed as representations of articulated meaning, encompassing both tangible and intangible phenomena \cite{36chandra2021levels}.

We apply Peirce’s components—representament, object, and interpretant—throughout our discussion and introduce the "cloud of context" as an additional design consideration for symbol-based information exchange around invisible disabilities.

\section{Related Work}

\subsection{Accessibility symbols and representation}

Accessibility symbols have long operated as both infrastructural wayfinding tools and public-facing representations of disability. Yet the most widely recognized marker—the International Symbol of Access (ISA)—has been repeatedly criticized for privileging mobility impairment and for encouraging narrow, medicalized interpretations of disability. Empirical evidence supports these critiques. Vice et al. (2020) show that participants with mobility impairments rated the ISA significantly more favorably than participants with non-mobility impairments (p < 0.002), and qualitative themes tied the ISA primarily to “wheelchair accessibility,” “restricted use,” and physical access rather than universal inclusion \cite{vice2020effectiveness}. This mismatch is especially consequential for people with invisible disabilities, whose access needs can be real but are less likely to be acknowledged through mobility-centered imagery.

Historical work helps explain why these representational limits persist. Guffey traces how the ISA emerged in parallel with wheelchair technology and barrier-free architecture, and how successive iterations reflect ongoing tensions between simplicity/standardization and community-led, identity-affirming representation \cite{guffey2}. This history shows that symbol design is never neutral: institutional pressures toward legibility and uniformity can marginalize diverse disability experiences and flatten identity into a single recognizable form. Together, these perspectives highlight a foundational problem motivating our study: a single dominant symbol can become “universal” in circulation while remaining exclusionary in meaning, particularly for those whose disabilities are non-mobility-based or invisible.

\subsection{Disability disclosure and hidden conditions}

Disclosing an invisible disability is not simply a matter of “making needs visible”—it is a decision shaped by perceived risk, audience, and anticipated outcomes. Work on online health information disclosure demonstrates that disclosure intentions follow a cost–benefit calculus influenced by privacy, trust \cite{HowCanISignal}, and individual dispositions. Other studies show that privacy concern functions as a disutility enhancer while trust functions as a disutility reducer, and that perceived information sensitivity mediates how personal dispositions (e.g., traits, perceived health status, prior privacy invasion) shape privacy concern and willingness to disclose \cite{bansal2010impact}. This framing aligns with how participants in our study describe selective visibility and hesitation: even when disclosure could support accommodation, uncertainty about who is watching and how information will be used can make disclosure feel risky.

Disclosure dynamics are also shaped by organizational settings and social influence. Extending a health-belief framing into workplace contexts, Various studies demonstrate that perceived susceptibility and consequences, perceived benefits, self-efficacy, and—critically—social influence predict participation in organizational online health programs \cite{nifadkar2021online, 25PinaInformatics, 2Workplace}. Their finding that coworker participation supports adoption parallels a key challenge for accessibility symbols: symbol uptake often depends on whether the practice is socially legible and normalized in the environment. Taken together, these studies emphasize that disclosure is context-sensitive, socially situated, and shaped by perceived interpretability—conditions that become especially complex when disclosure is mediated by symbols that may or may not be recognized.

\subsection{Symbol-enabled disclosure technologies and online engagement}

A growing body of work examines how digital environments shape the reception and circulation of health-related information \cite{towardpatient,designempathysocialmedia}, including how message form influences engagement and interpretation. Previous studies \cite{rus2016health} show that on diabetes-related Facebook pages, imagery is the strongest predictor of likes and shares, and that certain content features (e.g., consequence information, positive identity cues) predict sharing, while social support and crowdsourcing predict commenting. Importantly for symbol design, their findings suggest that visual form (image vs. text) does not merely “decorate” information—it shapes how audiences engage with it, and imagery can attenuate the effects of accompanying text. This is directly relevant to accessibility symbols and “carriers”: if a symbol is meant to enable disclosure, then the form factor and layering of information (glanceable icon, optional text, scannable link, or interactive element) can shape whether disclosure invites support, misunderstanding, or stigma.
In HCI, creative and speculative design traditions \cite{14Abowdstoryboarding} further show how future-oriented methods can surface values, risks, and desired safeguards—especially when current practices are constrained by stigma or limited infrastructures \cite{speculativemoreThanHuman}. Prior work on speculative and participatory methods has demonstrated the value of using artifacts such as sketches, scenarios, probes \cite{10SketchingDIS,11van2005sketching,12verstijnen1998sketching,13plurality} to elicit not only what people do today, but what they want to become possible—and what they fear new systems could enable \cite{designempathysocialmedia, speculativemoreThanHuman, speculativecodesign}. This motivates our approach: rather than treating accessibility symbols as static signage, we examine symbol-enabled disclosure as a sociotechnical system shaped by platform norms, carriers, audiences, and interpretive practices.

\subsection{Peirce-Semiotics and interpretation in symbol-based disclosure}

To theorize why symbol-based disclosure can succeed or break down, we draw on semiotics—specifically Peirce’s triadic account of meaning-making. In Peirce’s model, the \textit{representamen} is the perceivable form of the sign, the \textit{object} is what the sign refers to, and the \textit{interpretant} is the meaning produced by an interpreter \cite{22irvine2023semiotics}. This framing is useful for disability disclosure because misinterpretation is not incidental: it is a predictable outcome when observers infer beyond what a symbol can reliably communicate. In our context, the disability and access needs are not automatically “contained” in a symbol; they are constructed in interpretation, often under conditions of uncertainty. We build on Peirce’s triad by foregrounding the role of situational conditions that emerged in our findings. 
%

Although prior research establishes (1) the representational limitations of dominant accessibility symbols \cite{vice2020effectiveness, guffey2}, (2) the dispositional and contextual factors shaping disclosure decisions \cite{bansal2010impact, nifadkar2021online}, and (3) the importance of visual form in health-related communication and engagement online \cite{rus2016health}, these literatures rarely examine accessibility symbols as part of a broader disclosure system—where meaning depends on the symbol, the carrier through which it is surfaced, and the context that shapes interpretation. Our study addresses this gap by empirically examining how people with invisible disabilities reimagine “next-gen” accessibility symbols when paired with emerging modalities (wearables, mobile interfaces, portable devices, XR), and by using a semiotic lens to explain how misinterpretation and interpretant drift arise when visibility and context are not user-controlled.

\section{Methods}
Because our goal was to understand how people with invisible disabilities envision next-generation accessibility symbols working in future sociotechnical settings, we used a speculative co-creation approach grounded in established creative practices in HCI and design \cite{speculativemoreThanHuman, speculativecodesign} \cite{10SketchingDIS,11van2005sketching,12verstijnen1998sketching,13plurality}. Rather than treating accessibility symbols as static signage, we prompted participants to imagine symbols as part of a broader disclosure system. Our vision was for the participants to be able to focus on meaning that depends not only on the symbol itself, but also on the “carrier” that displays it, e.g., a wearable, a mobile interface, a portable device, or an XR overlay. This future-oriented framing was intentional. It enabled participants to surface values, risks, and safeguards related to visibility and interpretation that are difficult to elicit from accounts of current practices alone \cite{10SketchingDIS,11van2005sketching,12verstijnen1998sketching,13plurality}.

\textbf{Participants and Sample:} We recruited 23 participants who self-identified as having an invisible disability and completed an intake survey reporting disability identity, gender, and other demographics. Our sample included: ADHD (n=3; 2F, 1M), Autism (n=4; 3F, 1M), Blind/Low Vision (n=4; 2F, 2M), Diabetes (n=3; 3F), Dyslexia (n=4; 1F, 1M, 2NB), Epilepsy (n=1; 1M), Fibromyalgia Syndrome (n=2; 1F, 1M), and Parkinson’s (n=2; 1F, 1M). Participants’ lived experiences positioned them as experts of disclosure practices and interpretation risks, and our sessions were structured so that their artifacts and reflections drove the conversations. 

\textbf{Pre-session creative assignment (primary artifact data):}To center participants’ perspectives and provide multiple modes of expression, we emailed a pre-session activity one week before each session. Participants were asked to spend approximately 30 minutes (1) reflecting on accessibility symbols they encounter or use in daily life and (2) imagining 20 years into the future what it might look like to use technology to support symbol-based information sharing and disclosure. Participants responded through sketches. Blind and low-vision participants responded through creative writing. These artifacts were not treated as supplemental illustrations; they served as primary study data and the starting point for the co-creation session. We chose this format because visual and written artifacts can surface tacit needs, values, and concerns that may not emerge through interview prompts alone, and because visualizations can reveal needs that may be difficult to articulate verbally \cite{13plurality,14Abowdstoryboarding,15soya,16EvolvingPD}.

\textbf{Remote co-creation sessions and storyboards: } Each participant then completed a single 90-minute remote co-creation session. The first part of the session focused on the participants’ pre-session artifacts: participants explained the intent behind their sketches, the contexts they imagined, what they wanted to communicate, and what risks they anticipated (e.g., unwanted visibility, stereotyping, or misinterpretation). The second part introduced a set of semi-structured interview questions based on researcher-created storyboards as speculative probes—explicitly framed as conversation starters rather than proposed solutions \cite{speculativemoreThanHuman, speculativecodesign} [10–13]. These probes helped participants refine, critique, and extend their ideas, including articulating what they \textit{did not want} (e.g., “automatic labeling,” forced disclosure, or systems that enable othering). For blind and low-vision participants, we provided detailed written descriptions of each storyboard scenario to support equitable participation and accessibility in the co-creation process \cite{14Abowdstoryboarding,16EvolvingPD,17bodker}.

\textbf{Storyboards as speculative probes:} We developed seven storyboards (See Appendix~\ref{app:storyboards}) that build on recognizable symbol practices (badges, cards, lanyards) while extending them into plausible near-future carriers (wearables, interfaces, AR overlays). Each storyboard was designed to surface tensions central to symbol-based disclosure: selective visibility, audience, interpretation, and control \cite{11van2005sketching,12verstijnen1998sketching,14Abowdstoryboarding}.
\begin{enumerate}
    \item \textbf{Tap-to-share contact cards:} A phone-based exchange (similar to contact-card sharing) where a user can optionally display an accessibility symbol and selectively share short guidance about communication preferences or accommodations.
    \item \textbf{Social media badge:} A platform badge that can be toggled on/off to signal disability identity or access needs depending on context and audience.
    \item \textbf{Virtual meeting indicators:} A symbol displayed on a meeting tile (e.g., in a virtual meeting) to support access features such as captions and to enable selective disclosure in professional or educational settings.
    \item \textbf{“Just a Minute” cue in public spaces:} A wearable cue inspired by existing JAM cards to request patience in crowded environments, used to probe disclosure under time pressure and social scrutiny 
    \item \textbf{XR overlays (AR/VR/MR):} Smart-glasses overlays that surface accessibility cues in the environment
    \item \textbf{Everyday objects as carriers (e.g., payment cards):} A point-of-service scenario where an everyday object displays a symbol, prompting accommodations while raising questions about consent, visibility, and privacy.
\end{enumerate}

\begin{figure}[H]
\centering

\begin{subfigure}{0.49\textwidth}
  \includegraphics[width=\linewidth]{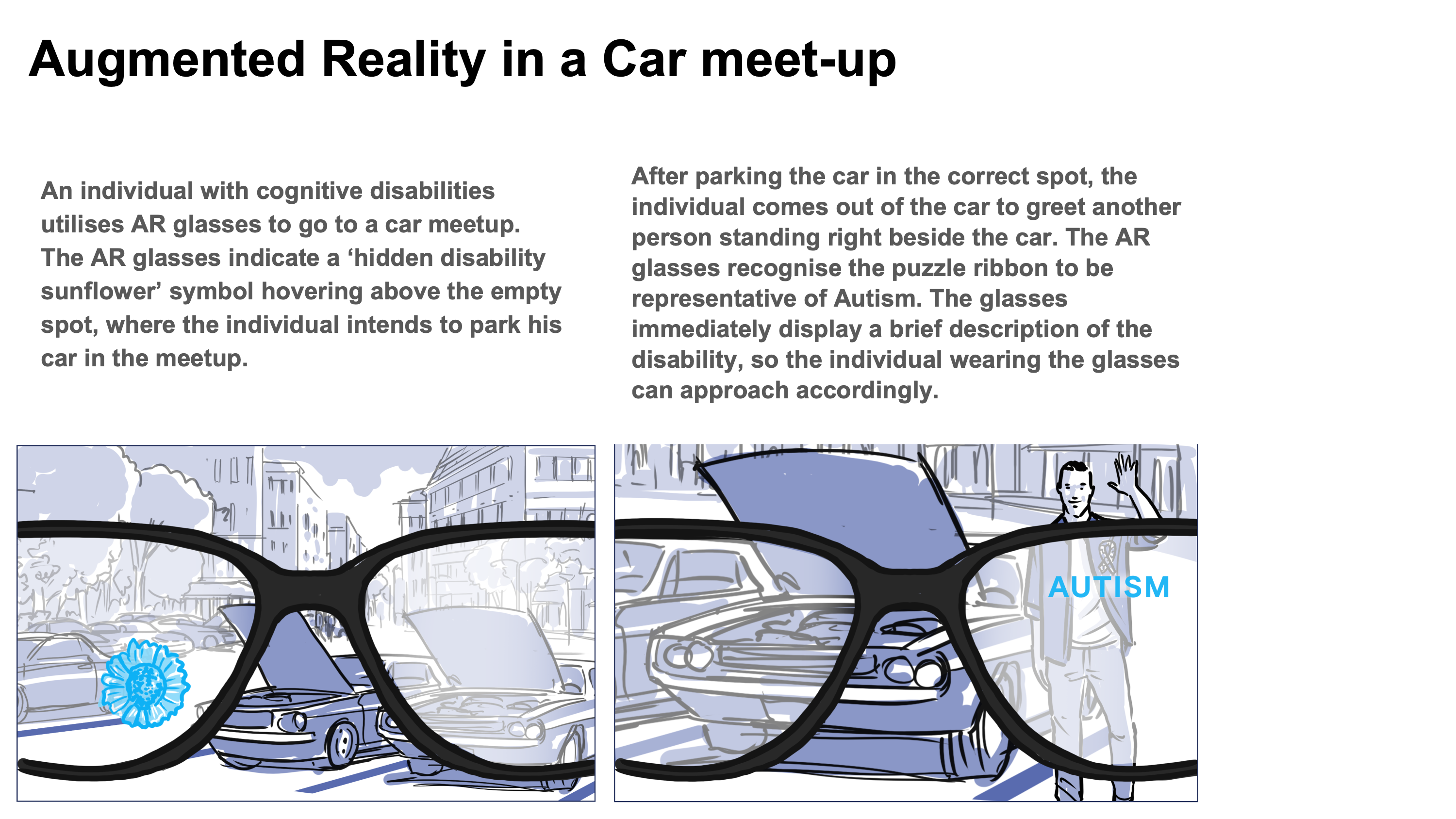}
  \caption{Storyboard 5}
\end{subfigure}\hfill
\begin{subfigure}{0.49\textwidth}
  \includegraphics[width=\linewidth]{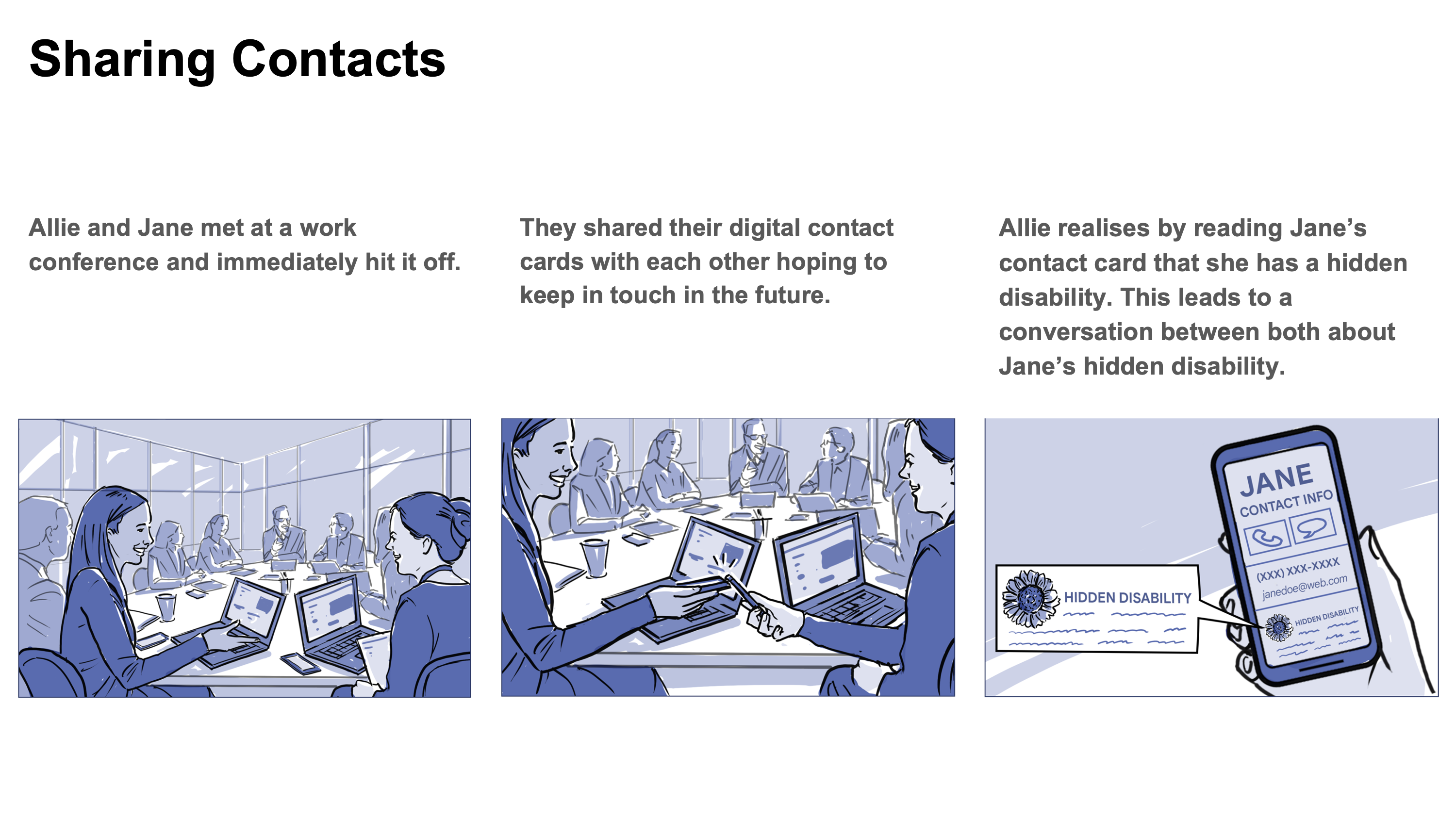}
  \caption{Storyboard 1}
\end{subfigure}


\end{figure}

Please note the storyboards' images use fictional names and scenarios for illustrative and rapport-building purposes only. Any resemblance to real persons is unintentional and purely coincidental. These creative artifacts informed the findings. We analyzed the recordings and transcripts using a crystallization–immersion approach \cite{borkan2022immersion, crabtree2022doing}. This method involves a horizontal strategy of interpretation characterized by multiple close readings of each transcript while remaining attentive to the lived experiences embedded in the text and open to emergent insights. As Crabtree and Miller note, “a single immersion often yields only superficial findings; repeated engagement with the data is necessary to capture depth and nuance” \cite{crabtree1992doing, crabtree2022doing}. Through this iterative process, we developed a richly contextualized understanding of participants’ orientations toward online health disclosure and the ways these shaped their broader social disclosures.

Across sessions, participant sketches and storyboard critiques were analyzed as situated proposals for how symbols might function in future disclosure systems using the immersion technique \cite{crabtree2022doing, borkan2022immersion}. We treated each artifact as evidence of the participant’s desired relationship between (a) symbol form and identity, (b) the carrier through which the symbol is displayed, and (c) the interpretability and risks produced in specific contexts. This approach allowed us to identify recurring patterns about when symbols are empowering versus risky, what controls participants require for selective disclosure, and how symbol carriers (from wearables to portable devices) change what can be communicated—from glanceable identity cues to more explorable, educational forms of disclosure \cite{14Abowdstoryboarding,16EvolvingPD,17bodker}.

\section{Findings}

Participants were asked to imagine symbol use 20 years into the future and to respond through sketches, creative writing, and discussion of storyboard probes. As a result, our findings reflect anticipated opportunities and risks—what participants want symbol-enabled disclosure to become, and what they fear it could enable—rather than a comprehensive account of current symbol practices.

The initial focus in our co-design sessions on symbol usage evolved into an exploration of how devices could integrate and translate symbols into communicative aids. Participants brought forward ideas for combining existing and future technology not just with symbols, but with customizable disclosure options that cater to diverse, representative preferences and needs. This revealed a complex interplay between symbols, meanings, and interpretations across various contexts.

Across these themes, participants repeatedly emphasized that symbols are rarely interpreted in isolation. Instead, the context in which a symbol is displayed and the form factor (or “carrier”) that conveys it (e.g., lanyards, bracelets, badges, mobile screens, portable devices, or XR overlays) shape what the symbol can communicate and how it is interpreted. Participants, therefore, treated symbols as part of a broader symbol-based disclosure system, where the same symbol can produce different interpretations depending on the audience, setting, and the degree of control the user has over visibility and accompanying information.

\subsection{Visibility, Interpretation, and Empowerment}
Participants emphasized the importance of making disability symbols visible across various contexts, including online spaces, workplaces, educational institutions, and virtual reality environments. Clear visibility was described as \textit{"empowering when it is controllable"}, and participants linked this to minimizing risks of misinterpretation or bias. They specified the importance of being able to control the visibility of the symbols and selectively share their disability symbol in contexts in which they want to share them.
For example, while enthusiastic about alternative symbols, P9 also expressed concerns regarding certain technologies. Regarding the use of augmented reality glasses to display disability symbols, they warned that \textit{"such applications could lead to harmful misinterpretations"} or \textit{“automatic labeling,”} posing risks of \textit{“othering,” “stereotyping,” or bias}. This highlights the potential for misuse of technology as a disclosure-communication aid if it lacks safeguards for context and individual intent.

\textbf{Semiotic finding for visibility and harm:} In Peirce’s terms, the accessibility symbol and its display mechanism function as the representamen (the perceivable sign). The disability and associated access needs form the object. P9’s concern foregrounds the interpretant—the meaning the observer constructs—which may become “automatic labeling” when the symbol is surfaced without consent or contextual grounding. This points to a core breakdown in symbol-based disclosure: interpretants can diverge sharply from the discloser’s intent when visibility is not user-controlled.

Building on the theme of visibility and interpretation, another shortcoming of symbols was the meaning behind them not being co-created with the community. As pointed out by P1, \textit{“a big thing in the disability community is making sure the disability community is driving the decision, right? I would want to talk to the community.”} P1 shared the \textit{“example I can think of is the puzzle piece for autism. Some people have really had strong opinions about it because it was developed by parents, but not by the community. So I want to make sure whatever symbol is adopted is like something the community wants, and they developed it themselves.”} We found that the symbols participants desired to use for representing hidden disabilities did not always originate from, or align with, their community—raising concerns about legitimacy, representation, and misinterpretation.

Participants also emphasized the need to redesign symbols, noting that the current ones failed to convey \textit{“a sense of empowerment.”}[P1, P9] They found this particularly important because, in their experience, non-disabled viewers often misinterpret disabilities as weaknesses. P1 shared they would want a symbol that looks \textit{“cool and more superhero-like”} than a disability. P6 suggested they would prefer \textit{“the ISA symbol but with a cape.”} These comments—often accompanied by sketches imagining more affirming aesthetics—illustrate that participants did not only want improved recognition; they wanted symbols that support identity and agency in disclosure moments. There is a generalization of symbols for the public, but then there are participants desiring to have a symbol that is their own personal representation. This representation changes in different contexts and with different people that they come across.

\textbf{Semiotic finding for Empowerment as interpretant: }Participants’ push for \textit{“superhero-like”} aesthetics highlights that the representamen (visual form) is not neutral: it shapes the interpretant by cueing strength, pride, or capability rather than deficit. In these accounts, empowerment is not simply a property of having a symbol; it emerges when the representamen supports a desired interpretation while still pointing to the object (the disability \&access needs) in a way the discloser controls.

Taken together, participants’ accounts show how empowerment is shaped by the interaction between (a) the symbol’s form and \textit{“feel,”} (b) \textit{"who is viewing it"} [P14], and (c) the user’s ability to control \textit{"when it becomes visible"}. P9 emphasized the need for controllable visibility to prevent misinterpretation and bias, and P1 built upon this idea by highlighting the need for these symbols to be co-created with the community, so that symbols are not only empowering but also reflect shared values and needs.

\subsection{Wearables as Symbol Carriers}

Our study’s participants expressed a desire for symbols that go beyond the generic, allowing them to tailor the information they disclose. Participants envisioned wearable devices that could adapt to different contexts, offering not just symbolic disclosure but also practical tools for sharing specific details as needed. They visualized wearable devices that could offer additional details—such as specific assistance needed, situational considerations, or even the specific type of disability.

For some participants, such as P13, the process of disclosure began with the difficult task of initiating discussions about their disability. P13 shared that disclosing one’s disability is a complicated topic to bring up. In such difficult moments, they felt that symbols could act as \textit{“conversation starters”} if combined with subtleties of wearable accessories. \textit{“You're always [wondering], how do I start this conversation? And [in the future] you [could] look at this band or bracelet with a symbol and say, ‘Maybe you're familiar with it?’ I've thought about this for years. I don't know why we don't have this.”} This highlights the potential of accessorized symbols to ease the emotional burden of initiating disclosure.

\textbf{Semiotic finding for Wearables as carriers:} In these scenarios, the wearable becomes part of the representamen—not only the icon itself, but the form factor that makes it glanceable, socially legible, and available at the right moment. The object remains the disability and the accommodation needs, but participants used wearables to shape the interpretant: “conversation starter” reframes the sign from a static label into a socially actionable cue.

P2 suggested that the wrist accessories could have \textit{“charms of the symbols of all my disabilities,”} indicating a desire for an accessory that communicates the presence of multiple invisible disabilities while also representing the uniqueness of each through distinct symbols as charms. She explained that she wanted a consistent wallpaper screen that could feature \textit{“a customizable non-medical handicap sign,”} either generalized or \textit{"representative of her primary disability"}, depending on what she chose to display. P7 shared a similar sentiment: \textit{“A bangle bracelet that is easy to put on and has the power to calm the tremor in my hand and arm. The decoration on it could say, ‘I have Parkinson’s.’”} This shifts the focus of accessibility symbols from merely identifying a condition to supporting participants’ desired degree of disclosure—ranging from a glanceable symbol to a more descriptive message—depending on context. As P15 shared: \textit{“[I would like to have a] wristwatch with a projector - what would be projected is information about some of my disabilities. For example, if you're having a hard time understanding me, please just ask me to repeat myself or, if you're having a hard time hearing me, let's try to move to a quieter location, things like that.”}
Participants emphasized the importance of merging symbolism with technology to enhance functionality, inclusivity, and empowerment. P22 reimagined the existing Hidden Disabilities Sunflower lanyard by suggesting \textit{“integrated tech capabilities.”} They described a scenario at the airport where wearing such a lanyard would signal a hidden disability, but they envisioned a more advanced design ––\textit{“sunflower lanyard with tactile features, perhaps Braille, or even a bump dot”} –– to help blind users ensure the lanyard is correctly oriented. They further suggested the inclusion of \textit{“QR codes… for airport staff to be able to see it and scan it, allowing for the provision of additional information or confirming its purpose.”} Taking the idea even further, they proposed, \textit{“I’d like to take it a step further and also have “a wearable device in the sunflower motif. It would not be on a wristband but around one’s neck. It would be as big as life because we are no longer ashamed of our disabilities but proud.”} This reflected a shift from discreet disclosure to a statement of pride and visibility, advocating for greater awareness and acceptance.

\textbf{Semiotic finding for Layered disclosure:} P22’s QR and tactile additions show participants imagining a representamen that can be layered—\textit{"a glanceable symbol"} for rapid recognition, plus optional paths to richer meaning (e.g., \textit{"scanning to learn more"}). This layering reflects participants’ desire to manage interpretants: allowing observers to confirm meaning rather than guess, while preserving user choice over how much detail is revealed.

\subsection {Portable Devices for Educational Sharing and Contextual Explanation}
Participant sketches further emphasize the importance of translating symbols onto adaptable, context-sensitive technologies that respect individual preferences and reduce the vulnerability often associated with visual disclosure. Importantly, participants did not treat portable devices as separate from symbols; rather, they imagined these devices as symbol carriers that can move from \textit{“glanceable” disclosure} (a symbol) to \textit{“explainable”} disclosure (additional details, examples, and tailored guidance). In contexts where symbols alone risk misunderstanding, participants wanted the ability to attach meaning, explanation, or accommodation guidance in ways that are still under the user’s control.

With participants as P17, disclosures were not just about their condition but also about \textit{“doing the work to educate and convince others”} about the severity and specificity of what they were experiencing. This part of the participants’ journey was not and could not be articulated by just having a symbol, neither ISA nor community-made symbols. Instead, participants imagined symbol-enabled systems where the symbol functions as an entry point—while the device provides optional, deeper contextualization when needed.

Examples of portable devices envisioned in sketches by participants resembled \textit{“a mini PC”} or \textit{“an iPad-like display”} (P3, P17, P4). These handheld devices could display vital information relevant to the user while also supporting disclosure by enabling participants to hand these devices out as needed.

P17, a participant with Type 1 Diabetes (T1D), shared her experiences with carrying a portable continuous glucose monitor. She noted how \textit{"the device often sparks curiosity"} among those who see her using it. Reflecting on her interactions, she explained: \textit{“When I disclosed T1D to someone, I’d hand them this device and let them mess around with it for a handful of minutes. Your average non-diabetic would be confused and overwhelmed by the dangerous art of diabetes management. They would get a glimpse into the mental labor, stress, and tediousness that characterize life with T1D and the constant threat of making a fatal error with insulin dosing. It’d be nice to get a little credit for living with all that.”}

\textbf{Semiotic finding for Device-as-sign, not drift:} P17’s example shows how a portable device can operate as a representamen that carries more than a label: the object is not only “T1D” in name, but also demonstrates the ongoing labor and risk of management. The interpretant emerges as the observer’s confusion and overwhelm, which becomes a route to recognition and empathy, supporting P17’s desire for disclosure to be understood as skilled work rather than a simple identity tag.

The vision for such portable devices that can be handed to whoever has questions about the disability comes from participants’ experiences of educating others about the complexity and cognitive fatigue involved in managing disabilities. By integrating features like QR codes that point to informational videos and educational content, participants looked to a future where these devices could foster greater understanding and empathy for the day-to-day challenges faced by individuals with chronic conditions—while allowing the individual to choose when education is appropriate versus when minimal disclosure is preferred.

Another example of this was P21, who also chose scenarios where he went from selective disclosure to educational disclosure: \textit{“If I am at a restaurant, the cashier or the waiter at the restaurant doesn't need to know necessarily that I'm autistic. But in my college courses, the students and faculty did not always have exposure to actual disabled people. So I started coming out as autistic, in most of my classes.. because it was really important to me that I know that my future colleagues have accurate exposure to disabilities.”}

P21 envisioned \textit{“a tiny portable printer that prints cards, and [the] technology would allow for selected details to be printed as wanted in the moment.”} They shared that they are currently using an app that allows them to pre-program messages about their disability and necessary accommodations so they don’t have to think on the spot in moments of stress or difficulty. This \textit{"pre-programmed information"} allows them to express their needs clearly and concisely \textit{"because it was prepared in advance"}. It allows them to educate the person on the intricacies of the disability while making it easier to advocate for accommodations. However, they have to include every possible accommodation they might need because they cannot easily edit or filter the information based on the context. They envision future technology that could allow them to selectively share only relevant details in the moment: \textit{“What I was imagining was, I could go into the app and check the boxes of the things that I am needing in this moment, press print and I have a card now that I can just give to the security desk or the restaurant manager or who whomever it is that I'm speaking to.”}

\textbf{Semiotic finding for Context selecting meaning:} P21’s account highlights how context determines what the sign should communicate: The same disability (object) does not require the same disclosure content in a restaurant versus a classroom. The portable printer becomes a mechanism for shaping the representamen in the moment (printing only what is relevant), thereby reducing interpretant drift (misunderstanding or oversharing) by aligning disclosure with the immediate situation.

Participants went beyond simply identifying and sharing their disabilities to tailoring the information needed within the immediate context to accommodate them. The context or situation in which the representation was happening was crucial. Their preference for portable devices to support interpretation—\textit{"especially in settings where misunderstandings are likely"}[P19, P21]—highlights their effort to prevent misinterpretations while preserving autonomy over what is shared, when it is shared, and with whom. This not only addressed immediate accommodation needs but also allowed participants to advocate for greater awareness and empathy by shifting from symbol-only disclosure to context-sensitive, optional explanation.

\section{Discussion}

Symbols alone often provide limited control over a condition’s narrative. Across our co-creation sessions, participants repeatedly treated symbols as entry points into disclosure rather than complete representations of disability. When paired with thoughtfully designed technologies—especially wearable and portable “carriers”—symbols became mechanisms for controlled disclosure, allowing individuals to balance privacy, autonomy, and legibility across contexts of information sharing. This aligns with a broader shift in assistive technology design away from generic, one-size-fits-all solutions toward more adaptive and personalized systems \cite{28Privacy}. In our study, personalization was not framed as “more information” by default; instead, participants emphasized selective visibility, context sensitivity, and interpretation management—including the ability to remain discreet in some settings and to invite explanation or education in others.

To explain why symbols sometimes succeed and sometimes break down in these scenarios, we elaborate Peirce’s Sign Complex model in semiotics \cite{19palinoan2024charles, 20carroll2023designers, 21mehawesh2014socio, 22irvine2023semiotics} as an analytic lens for disability disclosure. In Peirce’s model, a symbol involves three elements: the representamen (the perceivable form of the sign), the object (what the sign refers to), and the interpretant (the meaning produced in the observer’s mind). Our findings suggest that symbol-enabled disclosure depends not only on the symbol itself but also on the sociotechnical setting in which it appears. 

Peirce’s triad helps characterize what is being communicated and how interpretation is produced; yet, it offers limited guidance for when a particular interpretation becomes likely. For this reason, we add an overarching consideration we call the Cloud of Context: the situational, relational, and institutional conditions that shape how a symbol is displayed and interpreted. In our data, context was central to participants’ adoption and rejection of symbol-enabled systems, especially when participants anticipated risks of “automatic labeling” or misinterpretation in public or high-stakes environments \cite{18Gualano}.

Taken together, our findings answer the research questions as follows. Regarding RQ1 (Awareness), participants described uneven recognition across symbol types. While conventional accessibility signage is often legible at a glance, its meaning is frequently narrow, and condition-specific or emerging symbols can be personally resonant yet inconsistently recognized—making awareness a prerequisite for interpretability rather than a guarantee of effective disclosure [5, 31, 33]. For RQ2 (Usage), participants treated symbols as situational disclosure tools whose value depends on the ability to selectively reveal, hide, or elaborate meaning across settings (e.g., professional, educational, public, and online), consistent with prior work showing disclosure preferences shift with audience, stigma, and the affordances of the environment \cite{1ammari, 29joyammari, 30bruner2003chapter}. 

Addressing RQ3 (Interpretation breakdowns), participants anticipated misrecognition and “interpretant drift” when observers guess or oversimplify meaning, and they raised particular concern about technology-mediated visibility (e.g., XR) enabling “automatic labeling,” othering, or bias when context and consent are not respected \cite{18Gualano, 33handcock2004safety,37luborsky1994cultural}. For RQ4 (Reimagination with modalities), participants consistently reimagined symbols as part of layered disclosure systems in which the carrier (wearables, mobile interfaces, portable devices, XR overlays) becomes part of the sign itself, enabling movement from glanceable identity cues to optional, context-filtered explanation and education. Finally, for RQ5 (Motivations or Barriers), adoption and rejection of symbols were shaped by perceived empowerment, reduced disclosure labor, and legitimacy via community alignment, alongside barriers such as low awareness, fear of misinterpretation, and concerns about forced visibility or paternalistic symbol design.

Building on this synthesis of our research questions, in the subsections below, we use Peirce’s triad—situated within the Cloud of Context—to clarify how symbol-enabled disclosure succeeds or breaks down depending on what is represented (object), how it is surfaced (representamen), and how observers make meaning (interpretant)

\subsection{Representamen: from symbol-as-image to symbol-as-carrier}
Our findings underscore the importance of designing accessibility symbols and technologies that empower individuals while maintaining autonomy and control. Participants consistently emphasized the interplay between context, visibility, and representation in symbol-based communication. Prior research shows that individuals with stigmatized disabilities often prefer disclosure in contexts that afford anonymity \cite{1ammari, 29joyammari, 30bruner2003chapter}, which may be perceived as less judgmental than settings with known audiences. Consistent with this, participants such as P1 valued being able to control visibility in ways that reduce risks of “othering” or misunderstanding. Rather than suggesting that well-designed symbols eliminate stigma or discrimination, our data indicate a more specific effect: Participants anticipated that user-controlled visibility and clearer meaning could reduce vulnerability and misunderstanding in some disclosure situations, particularly when the symbol’s appearance and level of detail could be adjusted to the audience and setting.

A key contribution of our study is that participants treated the representamen as more than the graphic icon. In many examples, the carrier (e.g., charms on a bracelet, a lanyard, a projected watch display, a portable device, or an interface badge) was part of the representamen because it shaped visibility, readability, and the social meaning of disclosure. For example, participants’ desire for “conversation starters” and “charms of the symbols of all my disabilities” suggests that the wearable form factor is not merely a display surface; it helps make disclosure actionable and socially legible.

Peirce’s work is often used to distinguish symbols from signs that rely on resemblance, emphasizing that meaning is produced through interpretation rather than through direct visual similarity \cite{19palinoan2024charles, 20carroll2023designers, 21mehawesh2014socio, 22irvine2023semiotics}. In our context, this distinction matters: Participants did not require a symbol to “look like” a disability to represent it. This opens the design space beyond literal depictions such as wheelchairs, canes, or eyes. It supports participants’ interest in alternative aesthetics (e.g., “superhero-like” symbols) and community-driven forms that communicate identity without relying on body-part iconography. However, participants also noted that legitimacy and uptake depend on whether a symbol’s meaning aligns with how the community understands and wants to represent the condition—illustrated by P1’s critique of symbols developed without community leadership. In practice, this suggests that representamen design is not only a graphic problem; it is also a governance problem—who defines the sign and how it becomes recognizable through use.

Participants’ visions also surfaced a pragmatic constraint: A unique symbol for every disability may be desirable for representation but difficult to support in physical environments due to visibility and comprehension limitations. Here, participants positioned technology as a way to scale representation. Symbols can be rendered dynamically across online platforms, AR/VR/MR environments, and real-world settings—while allowing users to toggle, layer, or selectively disclose depending on context. In these visions, adaptive systems do not replace symbols; they alter the representamen by changing how symbols are surfaced, combined, and made explorable.

\subsection{Object: specificity, nuance, and what “the symbol refers to”}
In our study, the object pertains to disabilities and access needs. Our findings indicate that participants wanted symbols to communicate more specific and accurate meanings than generalized disability signage allows. For instance, participants with ADHD preferred recognition as having a neurodevelopmental condition rather than being implicitly mapped onto a mobility-centered symbol such as the ISA wheelchair. This aligns with critiques that the ISA is widely interpreted as a general disability signifier despite its strong association with mobility impairments and its limitations for cognitive, emotional, or sensory disabilities \cite{guffey2, 31barstow2019examining}. Our findings point towards how people with disabilities may prefer symbols that are more “humanized” rather than overly abstract \cite{31barstow2019examining, 32dall2010frequency}, which is relevant here because participants’ visions ranged from abstract identity marks to more narrative, descriptive disclosures.
Importantly, participants’ proposals suggest that “the object” of disclosure is often broader than a diagnosis label. In several accounts, the object included the functional implications of the disability and the accommodations needed in the moment. This was especially visible in examples where participants wanted symbols to be paired with optional details (“please ask me to repeat myself,” “move to a quieter location,” “here are the accommodations I need right now”). In that sense, symbol-enabled systems were imagined not only to signify “I have X,” but to communicate “here is what X means here,” in a way that remains under the user’s control.

Community-legible baselines with personalizable layers is a design recommendation that comes through. Participants wanted symbols that feel legitimate and empowering, and they explicitly cautioned against externally imposed representations (e.g., P1’s autism puzzle-piece example). Our findings suggest designing symbol systems with (a) a community-legible baseline—a recognizable, shared representation that supports comprehension and reduces interpretant drift—and (b) personalizable layers that allow individuals to tailor disclosure to context (e.g., multiple charms for multiple disabilities, optional text, QR links, toggles, or “conversation starter” affordances). Community-based co-creation and evaluation can help identify likely misinterpretations and refine symbols for clarity, while preserving the individual’s right to control how they appear in specific contexts \cite{shneiderman2020human} \cite{10SketchingDIS, 11van2005sketching, 12verstijnen1998sketching, 13plurality}

\subsection{Interpretant: interpretation risk, awareness, and preventing drift}
When it comes to disability, the consequences of misinterpretation can be significant and far-reaching. Participants highlighted the importance of providing specific details or educational context to reduce misinterpretations, particularly when observers lack familiarity with a symbol or when the disability’s effects are not externally visible. Interpretation varies with personal experience, cultural background, and learned behaviors \cite{37luborsky1994cultural}, making consistent interpretants difficult to guarantee. This is why participants’ visions frequently combined symbols with optional explanation—through QR codes, portable devices, or pre-programmed messages—so that interpretation could be guided rather than guessed.
Awareness is one factor that shapes interpretants: Familiarity with symbols can improve comprehension, and prior studies show a link between symbol familiarity and accurate interpretation \cite{33handcock2004safety}. However, our data also shows that awareness alone is insufficient when the context invites inference beyond the user’s intent. P9’s concerns about AR “automatic labeling” are instructive here: Even a recognizable symbol can produce harmful interpretants if it is displayed inappropriately or without consent \cite{18Gualano}. This suggests that symbol-enabled systems must treat interpretation as a design problem with failure modes, not as a passive outcome.
Participants’ visions, therefore, point toward interpretability mechanisms that can scale across contexts: interactive layers (e.g., an AR overlay that reveals details only when the user permits it), portable explainers (e.g., a device that invites observers to understand the labor of T1D management), or context-filtered printouts that allow users to share only what matters in the moment. Such mechanisms do not replace the symbol; they help stabilize interpretants by offering a controlled path from recognition to understanding.

Participants’ visions also surfaced a need for design for interpretant risk i.e make misinterpretation a first-class failure mode. Participants’ strongest concerns were not only about recognition, but about harms that emerge when observers infer too much (e.g., “automatic labeling,” stereotyping, or othering). Designers should therefore treat misinterpretation as a core risk and build safeguards that support user autonomy across contexts. Examples suggested by our findings include: (a) selective visibility (who can see the symbol and when), (b) layered disclosure (a glanceable symbol with optional, user-controlled explanation), and (c) context-filtered outputs (e.g., pre-programmed messages or printed cards that share only what is relevant in the moment). Where cultural differences may influence interpretation, designers should empirically test symbol meanings, aesthetic choices (including color), and representations across audiences rather than assuming universality \cite{23Konstantakis, 24Sartini, 25PinaInformatics, 33handcock2004safety, 37luborsky1994cultural}.

\section{Limitations}

Our study has several limitations that shape how the findings should be interpreted. First, our work used a speculative co-creation framing, asking participants to envision symbol-enabled disclosure 20 years into the future. As a result, the findings capture participants’ anticipated opportunities, concerns, and desired safeguards, rather than evaluating deployed systems or measuring real-world outcomes. 

Second, our sample (n=23) spans multiple disability communities with uneven representation across conditions, which limits condition-specific claims and suggests the need for deeper follow-up work within particular communities. 

Third, sessions were conducted remotely and relied on participant comfort with sketching and future-oriented ideation; this may bias participation toward individuals with access to stable technology and familiarity with articulating design ideas in creative formats. Fourth, while we centered participant-created sketches and narratives as primary evidence, visual artifacts can be challenging to interpret without context; we mitigated this by having participants explain their sketches during sessions, but the artifacts remain situated and may not generalize to broader populations. Finally, symbol interpretation is culturally and situationally variable; our recommendations therefore emphasize context-sensitive design and call for future evaluation with broader audiences (including non-disabled interpreters) and across settings to better understand recognition, misinterpretation, and potential harms at scale \cite{33handcock2004safety}.

\section{Conclusion}
Integrating accessibility symbols into digital and physical interfaces can provide subtle yet impactful assistance for individuals with disabilities, enabling them to communicate needs with greater confidence. Our research highlights how thoughtful design can bridge the gap between awareness and action by supporting context-sensitive, user-controlled disclosure—particularly for individuals with invisible disabilities whose needs are frequently misunderstood or overlooked. Across our speculative co-creation sessions, participants envisioned accessibility symbols not only as static indicators but also as components of broader disclosure systems whose meaning depends on the symbol’s form and the technological “carriers” through which it appears (e.g., wearables, portable devices, and digital interfaces). These visions also emphasize that improving disclosure requires designing for interpretation and misinterpretation: Participants valued symbols that can be selectively revealed, optionally explained, and aligned with community-defined meanings.

Through this investigation of accessibility symbols and their translation into emerging and wearable technologies, we underscore the critical role of design in promoting inclusivity while protecting autonomy. By attending to context, control, and interpretability, this work contributes to accessible design discourse and lays a foundation for future research and evaluation of symbol-enabled systems that support participation, normalize accommodations, and reduce vulnerability across everyday settings.

\section{Acknowledgments}

\bibliographystyle{ACM-Reference-Format}

\bibliography{sample-base}

\section{Appendices}

\appendix

\section{Storyboard set used in the speculative co-creation session} \label{app:storyboards}

\begin{figure}
\centering

\begin{subfigure}{0.49\textwidth}
  \includegraphics[width=\linewidth]{Slide2.png}
  \caption{Storyboard 1}
\end{subfigure}\hfill
\begin{subfigure}{0.49\textwidth}
  \includegraphics[width=\linewidth]{Slide3.png}
  \caption{Storyboard 2}
\end{subfigure}

\vspace{0.4em}

\begin{subfigure}{0.49\textwidth}
  \includegraphics[width=\linewidth]{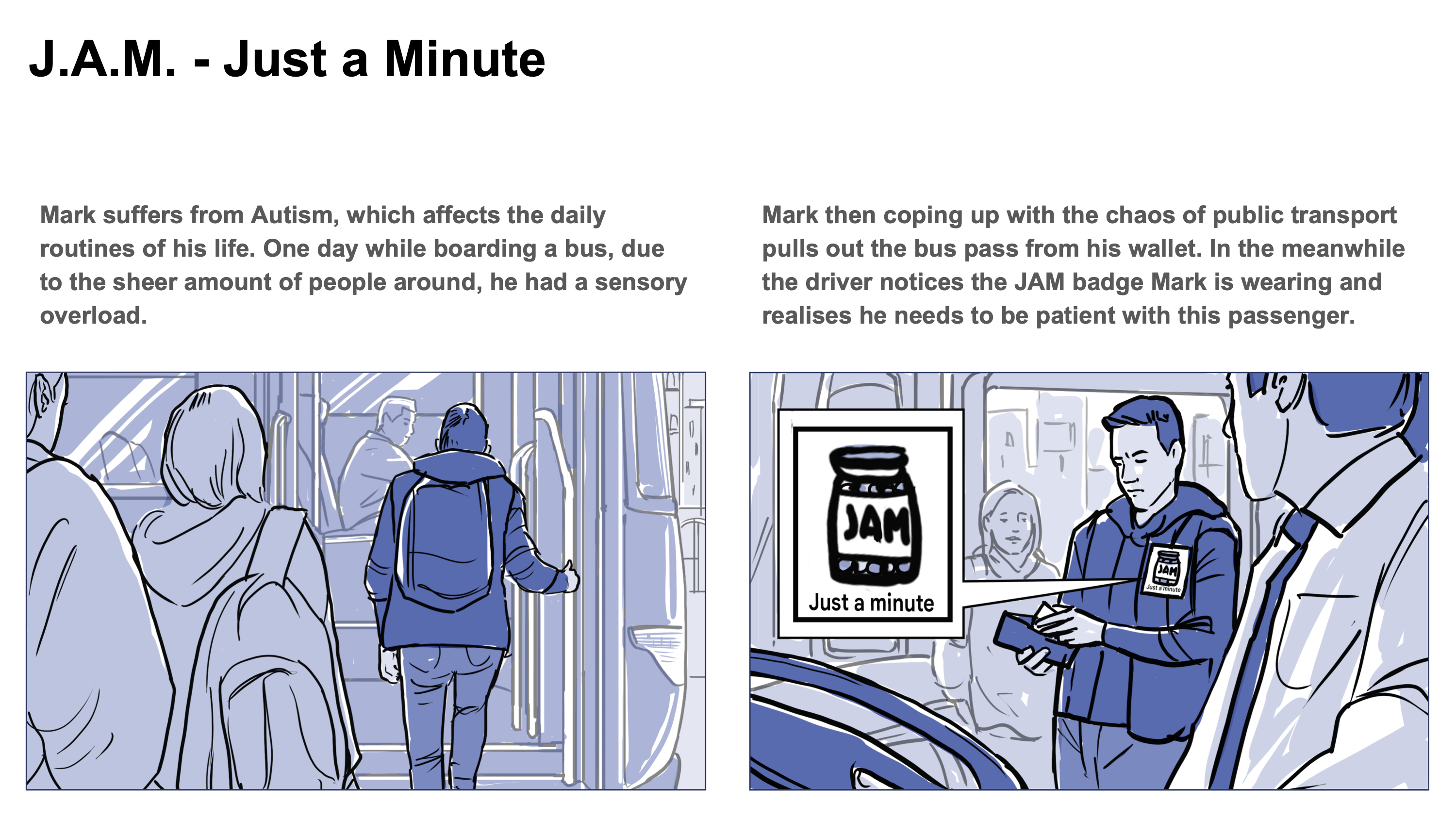}
  \caption{Storyboard 3}
\end{subfigure}\hfill
\begin{subfigure}{0.49\textwidth}
  \includegraphics[width=\linewidth]{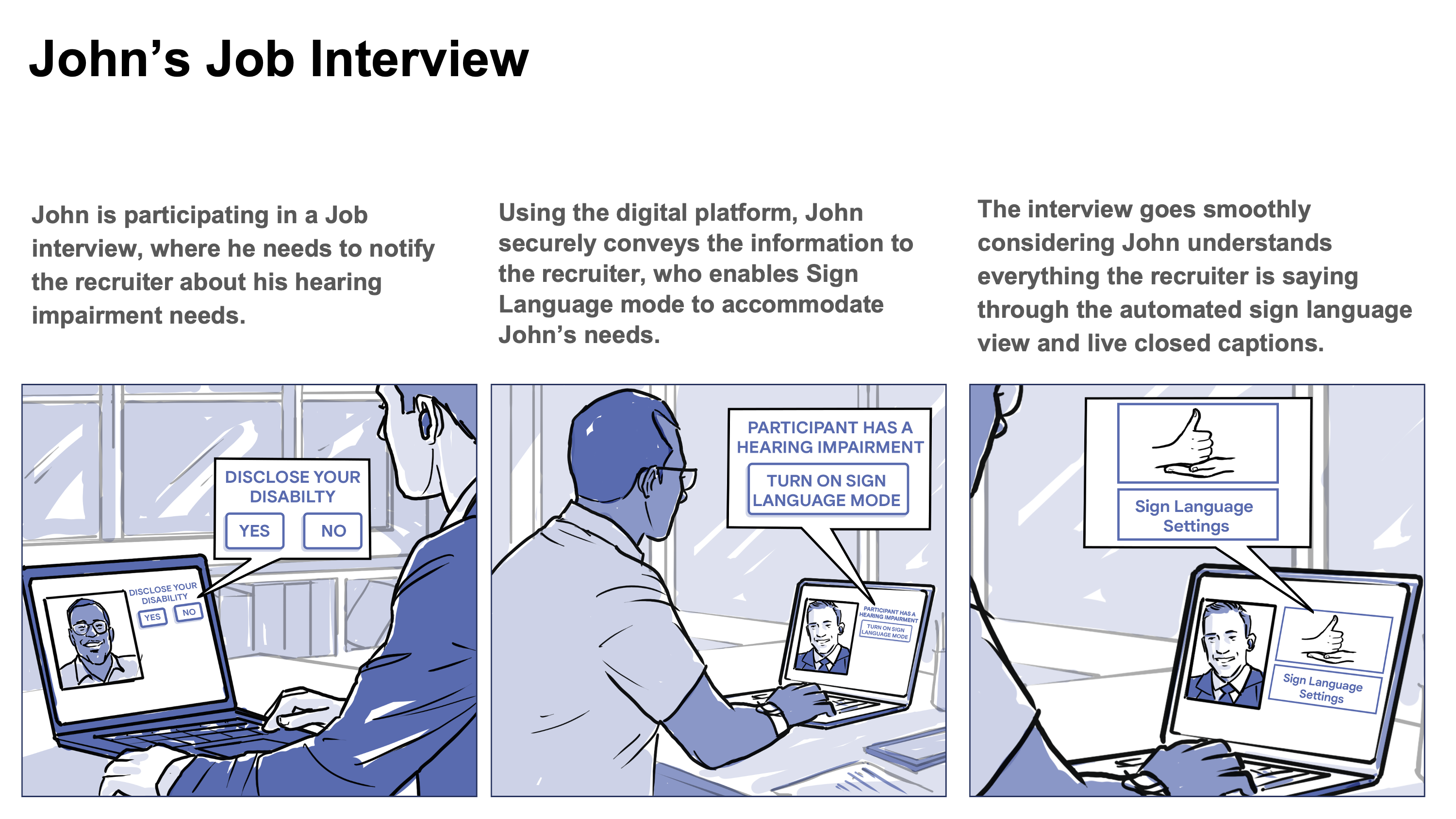}
  \caption{Storyboard 4}
\end{subfigure}


\centering

\begin{subfigure}{0.49\textwidth}
  \includegraphics[width=\linewidth]{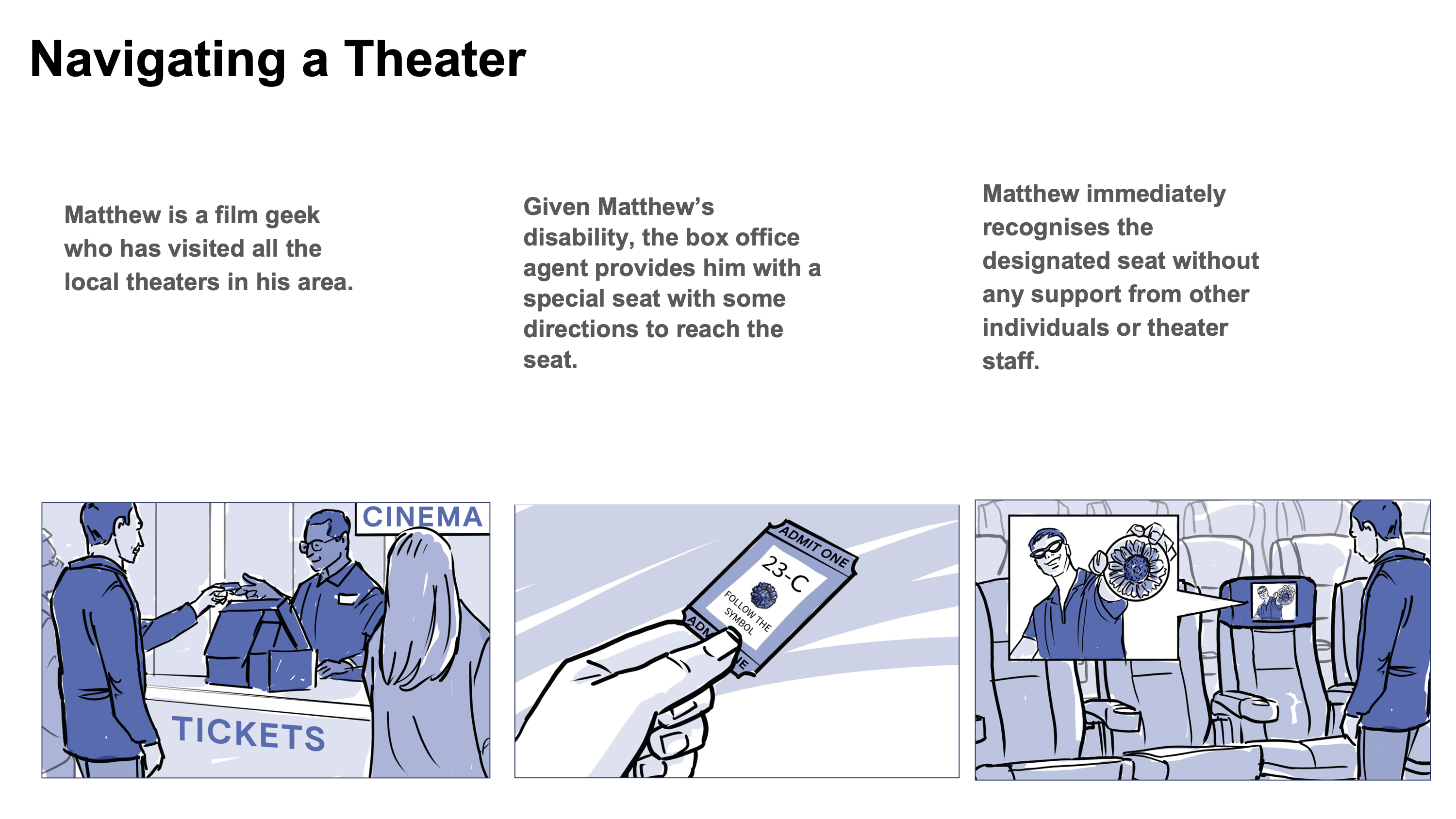}
  \caption{Storyboard 5}
\end{subfigure}\hfill
\begin{subfigure}{0.49\textwidth}
  \includegraphics[width=\linewidth]{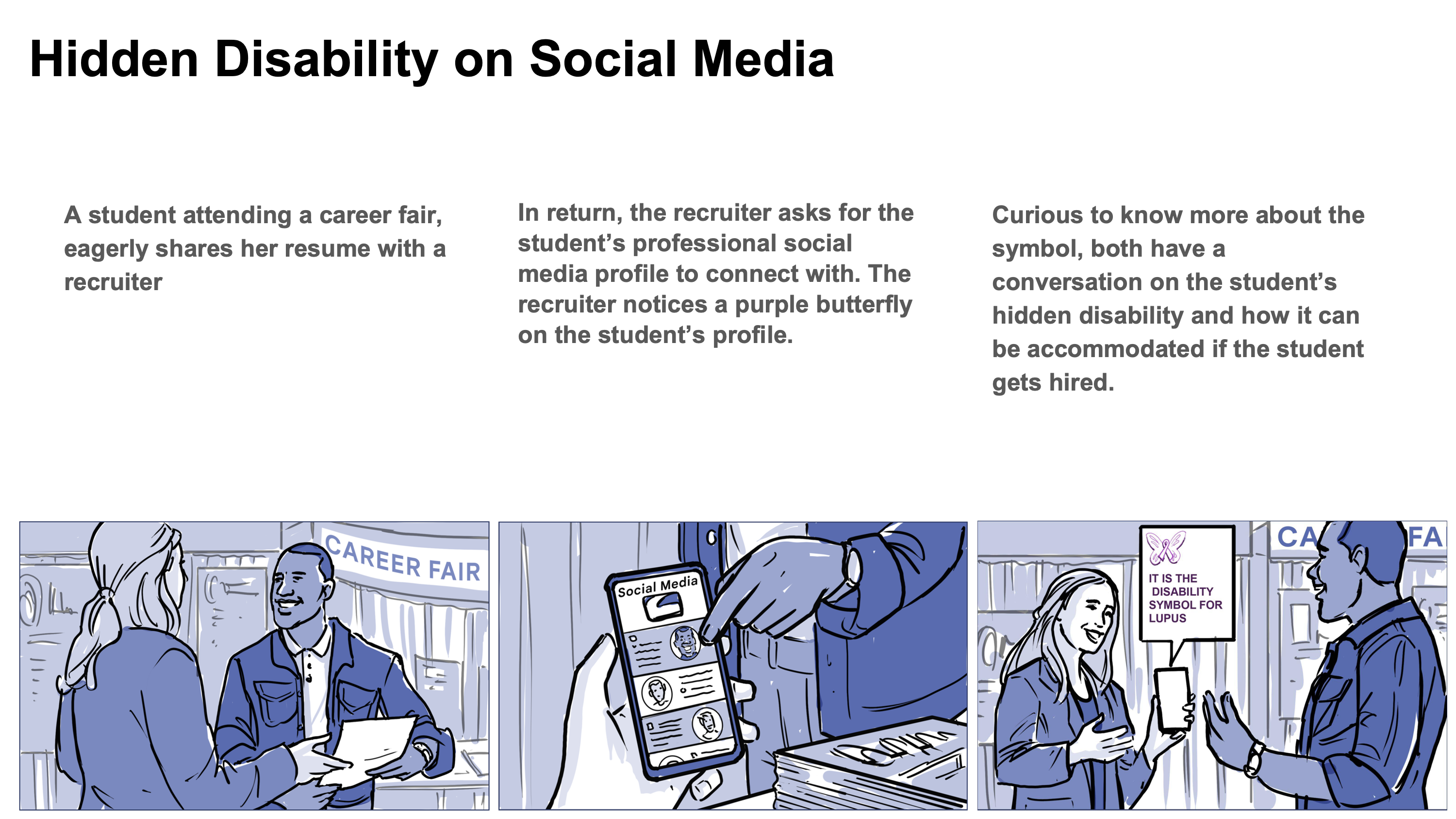}
  \caption{Storyboard 6}
\end{subfigure}

\vspace{0.5em}

\begin{subfigure}{0.70\textwidth}
  \centering
  \includegraphics[width=\linewidth]{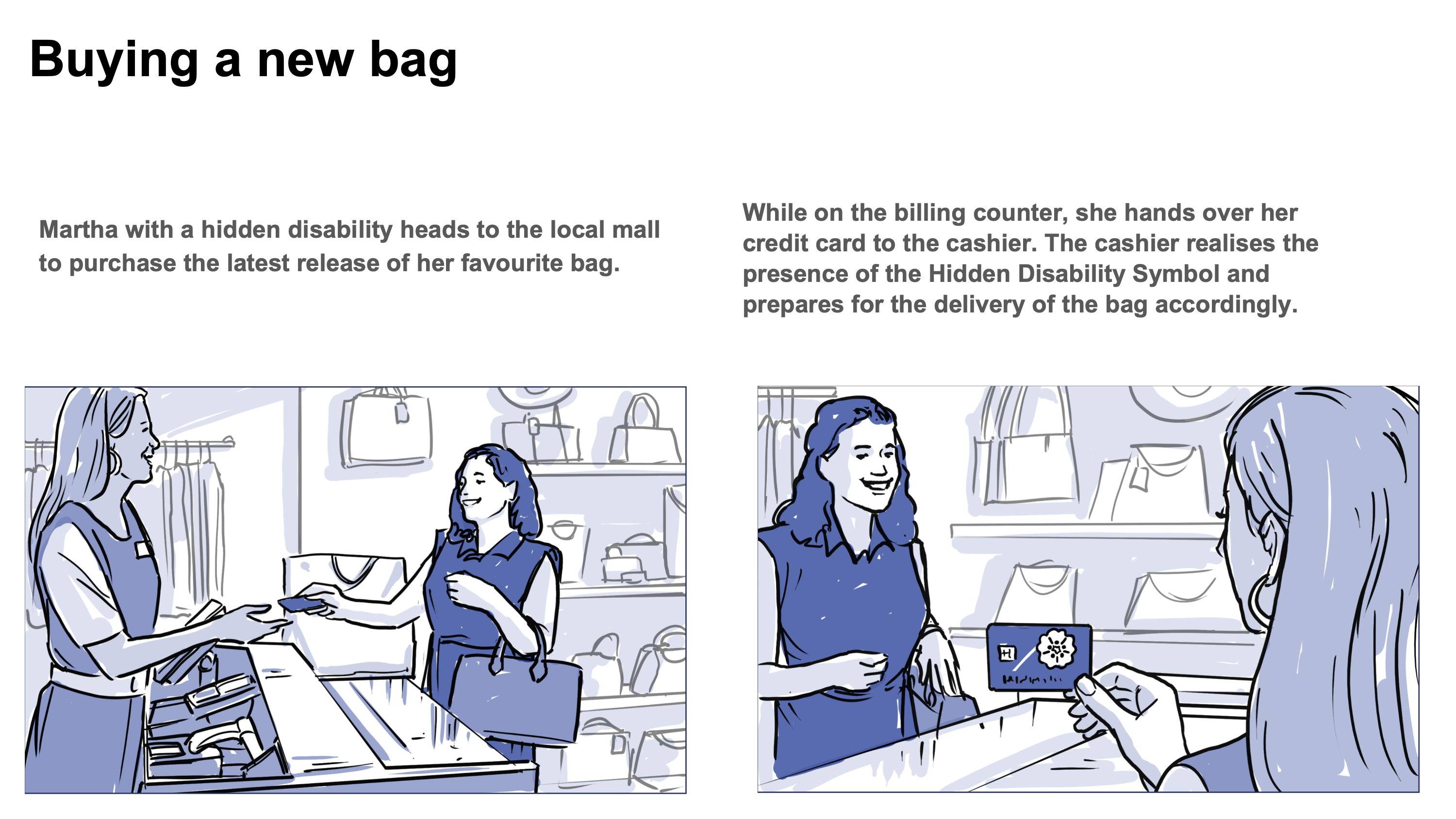}
  \caption{Storyboard 7}
\end{subfigure}

\end{figure}




\end{document}